\documentclass[aps,prl,reprint,superscriptaddress]{revtex4-1}

\usepackage{graphicx}
\usepackage{color}
\usepackage{textcomp}
\usepackage{siunitx}
\usepackage{xspace}
 \usepackage{mathtools}

\newcommand{\eg}{e.\,g.\xspace}

\begin{document}


\title{Electric-field dependent $g$-factor anisotropy in Ge-Si core-shell nanowire quantum dots}


\author{Matthias Brauns}
\email[Corresponding author, e-mail: ]{m.brauns@utwente.nl}
\author{Joost Ridderbos}
\affiliation{NanoElectronics Group, MESA+ Institute for Nanotechnology, University of Twente, P.O. Box 217, 7500 AE Enschede, The
Netherlands}
\author{Ang Li}
\email[Present address:]{Institute of Microstructure and Property of Advanced Materials, Beijing University of Technology, Pingleyuan No. 100, 100024, Beijing, People’s Republic of China.}
\affiliation{Department of Applied Physics, Eindhoven University of Technology, Postbox 513, 5600 MB Eindhoven, The Netherlands}
\author{Erik P. A. M. Bakkers}
\affiliation{Department of Applied Physics, Eindhoven University of Technology, Postbox 513, 5600 MB Eindhoven, The Netherlands}
\affiliation{QuTech and Kavli Institute of Nanoscience, Delft University of Technology, 2600 GA Delft, The Netherlands}
\author{Floris A. Zwanenburg}
\affiliation{NanoElectronics Group, MESA+ Institute for Nanotechnology, University of Twente, P.O. Box 217, 7500 AE Enschede, The
Netherlands}


\date{\today}

\begin{abstract}
We present angle-dependent measurements of the effective $g$-factor $g^\star$ in a Ge-Si core-shell nanowire quantum dot. $g^\star$ is found to be maximum when the magnetic field is pointing perpendicular to both the nanowire and the electric field induced by local gates. Alignment of the magnetic field with the electric field reduces $g^\star$ significantly. $g^\star$ is almost completely quenched when the magnetic field is aligned with the nanowire axis. These findings confirm recent calculations, where the obtained anisotropy is attributed to a Rashba-type spin-orbit interaction induced by heavy-hole light-hole mixing. In principle, this facilitates manipulation of spin-orbit qubits by means of a continuous high-frequency electric field.
\end{abstract}


\maketitle

\section{Introduction}
Quantum computation \citep{Aaronson2013quantum,DiVincenzo1995,Ladd2010} has made an enormous leap from a far-fetched promise \citep{Feynman1986a} to a realistic near-future technology \citep{Vandersypen2001,Jones2012,Veldhorst2015} during the past three decades. Among others, spin systems in the solid state \citep{Loss1998,Kane2000} have been developed into a mature but still very fast-evolving research field. In recent years increased research efforts have focused on C, Si, and Ge \citep{Laird2014,Zwanenburg2013,Amato2014}, which can be purified to only consist of isotopes with zero nuclear spin \citep{Itoh2003, Itoh1993} and thus exhibit exceptionally long spin lifetimes \citep{Muhonen2014a,Veldhorst2014}.
\paragraph{}
The one-dimensional character of electrostatically defined quantum dots in Ge-Si core-shell nanowires leads to unique electronic properties in the valence band, where  heavy and light hole states are mixed \citep{Csontos2007,Csontos2009,Kloeffel2011a}. The band mixing gives rise to an enhanced Rashba-type spin-orbit interaction \citep{Kloeffel2011a}, leading to strongly anisotropic and electric-field dependent $g$-factors \citep{Maier2013}. This makes quantum dots in Ge-Si core-shell nanowires promising candidates for robust spin-orbit qubits that can be electrically controlled via circuit quantum electrodynamics \citep{Kloeffel2013a}. 
\paragraph{}
Despite these profound theoretical contributions, only few experiments in Ge-Si core-shell nanowires have been reported including Josephson junctions \citep{Xiang2006b}, spin-filling \citep{Roddaro2008}, spin relaxation \citep{Hu2012}, spin coherence \citep{Higginbotham2014a},  charge sensing \citep{Hu2007} in the many-hole regime, and signatures of weak antilocalization \citep{Higginbotham2014}.
\paragraph{}
In this work we experimentally explore the {an-iso-tro-py} of the $g$-factor in Ge-Si core-shell nanowires. We electrostatically define a highly-tunable, elongated hole quantum dot in the nanowire by means of local gates. We measure the Zeeman splitting of a single-particle state in the quantum dot while rotating the magnetic field around the high-symmetry axes of the system and find a strong anisotropy with respect to the nanowire as well as to the electric field, in line with theoretical predictions \citep{Maier2013}.

\section{Gate-defined quantum dots}

\begin{figure}
\includegraphics{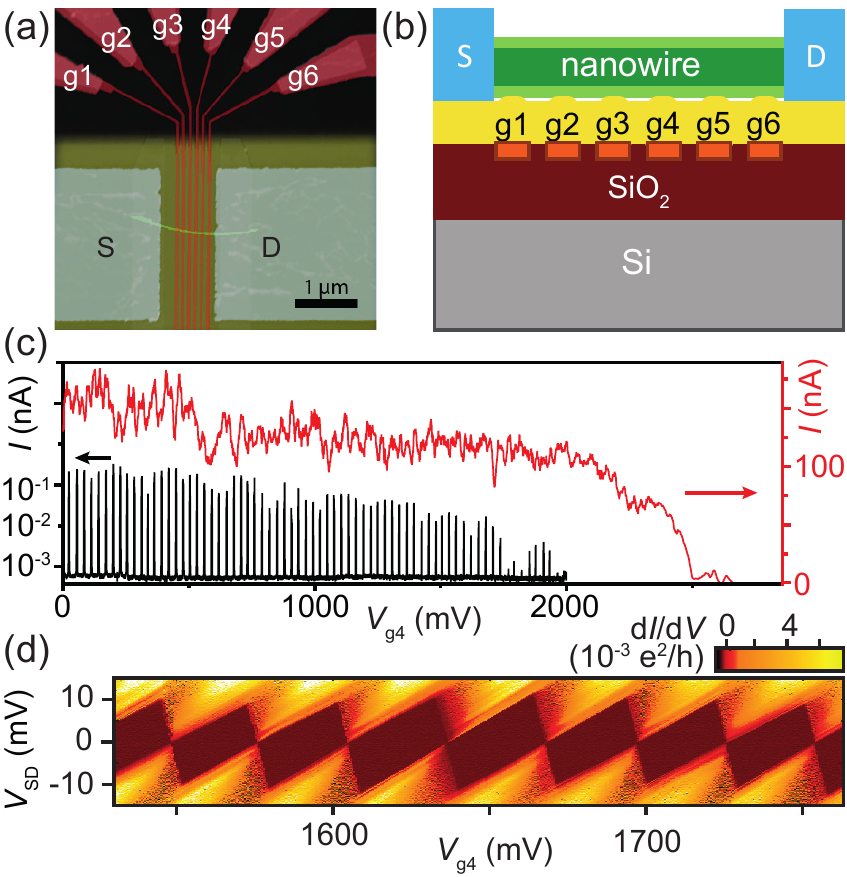}%
\caption{(a) False-color atomic-force microscopy image of the device. (b) Schematic cross-section displaying the p++-doped Si substrate (grey) with 200 nm of SiO$_{\text{2}}$ (dark red), six bottom gates g1-g6 (light red), each $\sim$35~nm wide and with 100~nm pitch buried under 10~nm Al$_{\text{2}}$O$_{\text{3}}$ (yellow), on top of which the nanowire is placed (green) with ohmic contacts (0.5/50~nm Ti/Pd, blue). (c) Current $I$ vs. $V_{\text{g4}}$ with g3 and g5 forming tunnel barriers ($V_{\text{g3}} = 2060$~mV, $V_{\text{g5}} = 2260$~mV). Black curve is taken at $V_{\text{SD}} = 1$~mV, red curve at $V_{\text{SD}} = 50$~mV. (d) Numerical differential conductance $dI/dV_{\text{SD}}$ plotted vs. $V_{\text{SD}}$ and $V_{\text{g4}}$ at the same barrier voltages as in (c).\label{Fig1}}
\end{figure}

Our device in Fig.~\ref{Fig1}(a) consists of a p$^{\text{++}}$-doped Si substrate covered with 200~nm SiO$_{\text{2}}$, on which six bottom gates with 100~nm pitch are patterned with electron-beam lithography (EBL). The gates are buried by 10~nm Al$_{\text{2}}$O$_{\text{3}}$ grown with atomic layer deposition at $100^{\circ}$C. A single nanowire  with a Si shell thickness of $\sim$2.5~nm and a defect-free Ge core with a radius of $\sim$8~nm \citep{Li2016} is deterministically placed on top of the gate structure with a micromanipulator and then contacted with ohmic contacts made of 0.5/50~nm Ti/Pd. A source-drain bias voltage $V_{\text{SD}}$ is applied to the source, the current $I$ is measured between the drain and ground. All measurements are performed using dc electronic equipment in a dilution refrigerator with a base temperature of 8~mK and an effective hole temperature of $T_{\text{hole}} \approx 30$~mK determined by measuring the temperature dependence of the Coulomb peak width \cite{Mueller2013,Goldhaber-Gordon1998}.

\paragraph{}
We use this gate design \cite{Fasth2005a} to electrostatically define a single quantum dot \cite{sohn1997mesoscopic,Fasth2005a}. The two barrier gates g3 and g5 control the tunnel barriers, and a third plunger gate g4 the electrochemical potential of the quantum dot. In Fig.~\ref{Fig1}(c) we plot $I$ versus the voltage on g4 $V_{\text{g4}}$. When applying a high $V_{\text{SD}} = 50$~mV we observe a strong suppression of $I$ for $V_{\text{g4}} > 2.5$~V, indicating depletion of the nanowire at  $V_{\text{g4}} \approx 2.5$~V. At low $V_{\text{SD}} = 1$~mV we observe Coulomb peaks \cite{sohn1997mesoscopic} with a regular spacing of $\Delta V_{\text{g4}} \approx 30$~mV over a range of 2~V, i.e. we are able to change the hole occupation of the quantum dot by more than 60 holes. Above $V_{\text{g4}} \approx 2$~V no regular Coulomb peaks are observed, but the high-bias gate sweep suggests that the quantum dot is not completely emptied, i.e. in this device we are unable to identify the last hole on the quantum dot. If we assume the plunger gate coupling to stay constant and the quantum dot to be empty at $V_{\text{g4}} = 2.5$~V, we can estimate the number of remaining holes to be $N \approx 17$ at $V_{\text{g4}} = 2$~V. Reaching the single-hole regime was not possible in our device.
\paragraph{}
A non-linear transport measurement is displayed in Fig.~\ref{Fig1}(d). In this bias spectroscopy we plot the numerical differential conductance $dI/dV \equiv dI/dV_{\text{SD}}$ vs. $V_{\text{SD}}$ and $V_{\text{g4}}$, as will be in all the following bias spectroscopy plots. Formation of a single quantum dot is indicated by regularly shaped, closing Coulomb diamonds \cite{sohn1997mesoscopic}. The height of the Coulomb diamonds indicates an addition energy of $E_{\text{add}} \approx 8 - 10$~meV. The variations in $E_{\text{add}}$ cannot be explained by an interacting second quantum dot, which would lead to non-closing diamonds. Orbital shell filling can cause the variations in $E_{\text{add}}$ \cite{Tarucha1996,Kouwenhoven2001}. The low number of residing holes ($\sim$ 25-30) supports this reasoning.
\paragraph{}
The results in Fig.~\ref{Fig1} show a highly tunable nanowire device in which we intentionally define a very stable quantum dot. We can control the number of holes in the quantum dot over a wide range from approximately 85 down to approximately 17.

\section{Zeeman splitting of the orbital ground state}

\begin{figure}
\includegraphics{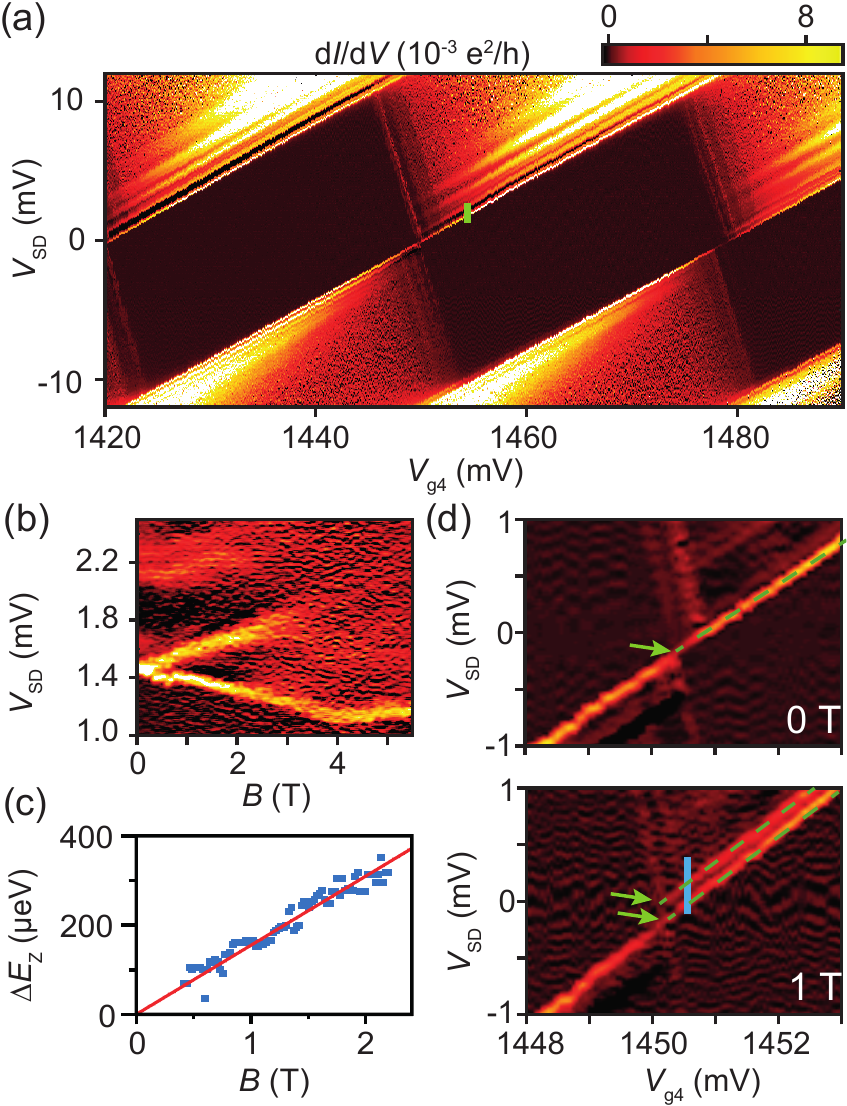}%
\caption{(a) Bias spectroscopy of the charge transition used for magnetospectroscopy measurements. The $dI/dV$ color scale applies to all bias spectroscopies in this figure. (b) $dI/dV$ vs. $V_{\text{SD}}$ and $B$ along the green line in (a). (c) $\Delta E_{\text{Z}}$ extracted from b) vs. $B$ together with a linear fit (red line) that yields $g^\star = 2.7\pm0.1$. (d) Bias spectroscopies measured at $B = 0$ (upper panel), and $B = 1$~T (lower panel). The green arrows indicate the spin-degenerate and spin-split orbital ground state. Measurements in Fig.~\ref{Fig3} were taken along the blue line.\label{Fig2}}
\end{figure}

We now investigate the Zeeman splitting $\Delta E_{\text{Z}}$ of the spin-degenerate quantum dot states \cite{Hanson2007}. To determine the $g$-factor with high accuracy, we choose a charge transition where the onset of conductance is sufficiently separated from other lines of increased conductance corresponding to, e.g., (orbital) excited states of the quantum dot, or resonances due to the low dimensionality of the leads \cite{Escott2010}. In Fig.~2(a) we show a bias spectroscopy of such a charge transition together with the two Coulomb diamonds adjacent to it. The number of residing holes here is approximately 35, again estimated by comparing the current plunger gate voltage ($V_{\text{g4}} \approx 1450$~mV) to the depletion voltage at high bias ($V_{\text{g4}} \approx 2.5-2.7$~V). 
\paragraph{}
We keep the plunger-gate voltage at $V_{\text{g4}} = 1454.0$~mV and sweep  $V_{\text{SD}}$ along the green line in Fig.~\ref{Fig2}(a) at different magnetic fields $B$ while measuring the current [Fig.~\ref{Fig2}(b)]. The magnetic field here is applied in the plane of the chip perpendicular to the nanowire axis. At $B = 0$, one very pronounced peak marks the onset of conductance, which splits up into a spin-ground and spin-excited state at finite magnetic fields. Note that the shifts of the two states are symmetric and linear up to at least 2~T, indicating that for magnetic fields $B < 2$~T the linear Zeeman splitting is the only relevant term, and other effects, such as a diamagnetic shift \cite{Rinaldi1996,Zielke2014}, are negligible. The spin splitting of the orbital ground state is further confirmed by the two bias spectroscopies in Fig.~\ref{Fig2}(d) at $B = 0$ and $B = 1$~T. The spin-degenerate orbital ground state of the charge transition at $B = 0$ [indicated by a single green arrow Fig.~\ref{Fig2}(d)] is clearly split into two lines at $B = 1$~T (indicated by two green arrows).
\paragraph{}
We extract the Zeeman splitting $\Delta E_{\text{Z}}$ by converting the $V_{\text{SD}}$ scale into energy. The lever arm $\alpha \equiv C_{\text{tot-S}} / C_{\text{tot}}$ with $C_{\text{tot-S}} = C_{\text{tot}}-C_{\text{S}}$, where $C_{\text{tot}}$ is the total capacitance of the dot, and $C_{\text{S}}$ is the source capacitance) for this conversion is graphically extracted: the slopes of the Coulomb diamond edges from Fig.~\ref{Fig2}(a) are $a \equiv |-C_{\text{G}} / (C-C_{\text{S}})| = 2.94$ and $b \equiv C_{\text{G}} / C_{\text{S}} = 0.44$, where $C_{\text{G}}$ is the gate capacitance \citep{Hanson2007}. By using $\alpha = 1 / (1 + b/a)$ we find a lever arm of $\alpha = 0.87$. The linear increase of $\Delta E_{\text{Z}}$ with increasing $B$ is shown in Fig.~\ref{Fig2}(c).  We fit the slope of $\Delta E_{\text{Z}}$ according to $\Delta E_{\text{Z}} = g^\star \mu_{\text{B}} B$, where $g^\star$ is the effective $g$-factor, and $\mu_{\text{B}}$ is the Bohr magneton [see Fig.~\ref{Fig2}(c)]. This yields an effective $g$-factor for this transition of $g^\star = 2.7\pm0.1$. We point out that the spin states are mixtures of heavy and light hole states and therefore $m_{\text{s}} \neq 1/2$, which is accounted for by the introduction of $g^\star$ as an effective $g$-factor.  Note that $g^\star$ may differ significantly from transition to transition due to the varying heavy-hole light-hole mixing of subbands and quantum dot states \cite{Kloeffel2011a} at the valence band edge of the nanowire \cite{Roddaro2008,Kloeffel2013a}.
\paragraph{}
In summary, in Fig.~\ref{Fig2} we determine the effective $g$-factor $g^\star$ to be $g^\star = 2.7\pm0.1$ for an in-plane magnetic field perpendicular to the nanowire. The corresponding Zeeman splitting is symmetric and linear up to at least 2~T.

\section{g-factor anisotropy}

\begin{figure*}
\includegraphics{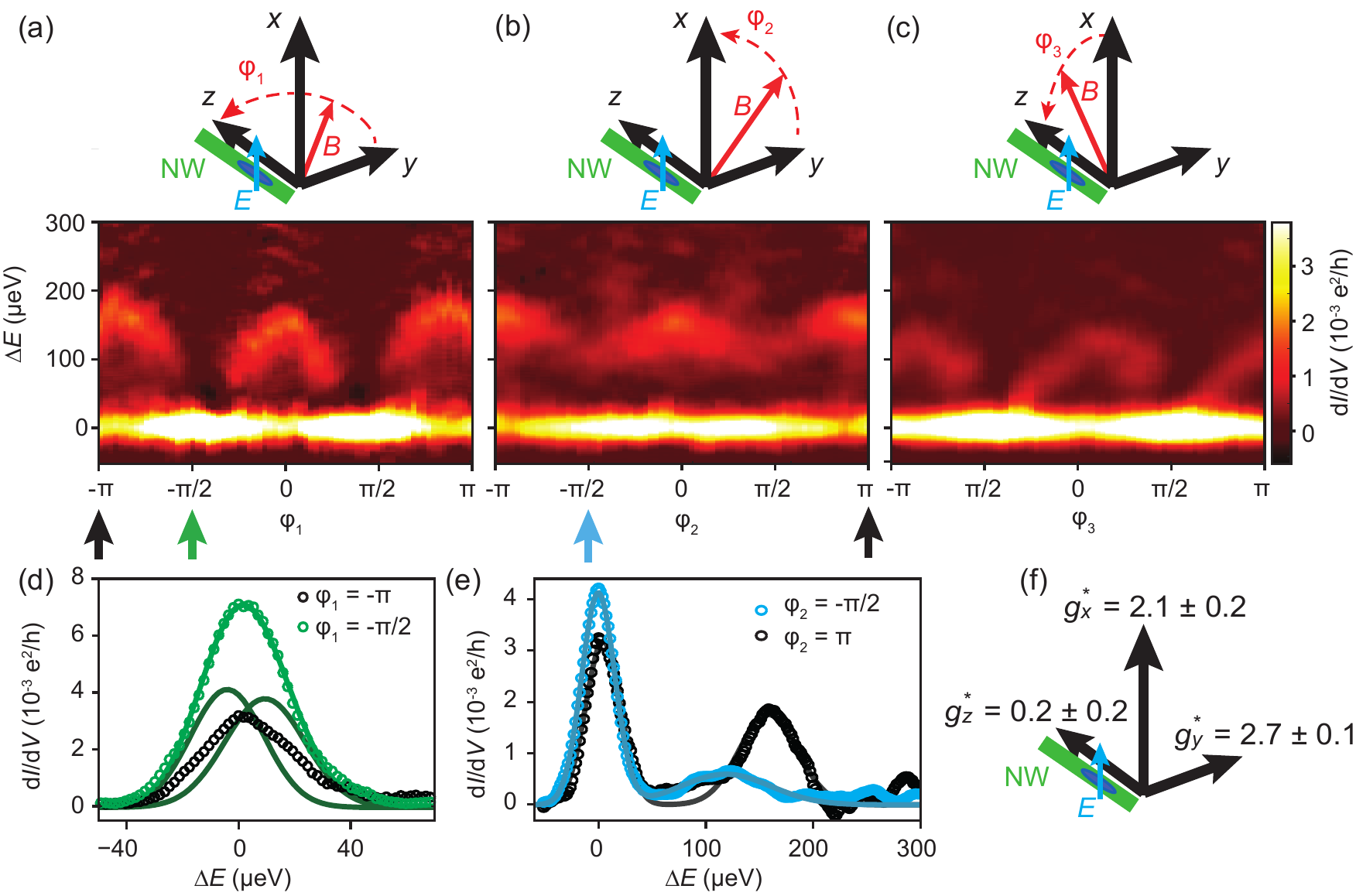}%
\caption{a) $dI/dV$ plotted for different $\vec{B}$-field directions at constant $B = 1$~T measured along the blue line in Fig.~\ref{Fig2}(d). Rotation of the magnetic field along (a)  $\phi_1$ in plane of the chip  (b) $\phi_2$ in the plane perpendicular to the nanowire axis. (c) $\phi_3$ from the electric-field axis to the nanowire axis. (d) line cuts taken from (a) at $\phi = -\pi$ (black circles) and $\phi = -\pi/2$ (green circles) plotted together with the fit (green line as the sum of the dark green lines) for the $\phi = -\pi/2$ line cut. (e) line cuts taken from (b) at $\phi = \pi$ (black circles) and $\phi = -\pi/2$ (blue circles) along with the respective fits (black and blue lines) (f) summary of the measured $g$-factors along the high-symmetry axes.\label{Fig3}}
\end{figure*}

To investigate the anisotropy of the $g$-factor, we measure the Zeeman splitting of the ground state at a fixed magnetic field magnitude of $B \equiv |\vec{B}| = 1$~T while changing the direction of $\vec{B}$. We choose the coordinate system in accordance with Maier \emph{et al.} \cite{Maier2013}, i.e. the $z$-axis points along the nanowire axis, the $x$-axis points out of the chip plane parallel to the electric field produced by the bottom gates, and the $y$-axis is in plane with the chip and perpendicular to the nanowire [see Fig 3(a)]. We will show measurements in three orthogonal rotation planes. Within each plane, a full $2 \pi$-rotation of $\vec{B}$ is performed in steps of $\pi/36$. For each step, $I$ is measured vs. $V_{\text{SD}}$ along the blue line in Fig.~\ref{Fig2}(d). The values for $\Delta E_{\text{Z}}$ along the different directions are obtained by fitting the line cuts with two peaks for the spin-ground and spin-excited states and calculating the distance between the two peak centers.
\paragraph{}
First we rotate the magnetic field from the $y$-axis to the $z$-axis [Fig.~\ref{Fig3}(a)]. At $\phi_1 = 0$, the Zeeman splitting of $\Delta E_{\text{Z},y} = 155 \pm 5$~$\mu$eV corresponds to a $g$-factor of $g^\star_{y} = 2.7\pm0.1$ [in agreement with Fig.~\ref{Fig2}(c)]. The Zeeman splitting decreases when the magnetic field is rotated towards the nanowire axis, until it is almost completely quenched at $\phi_1 = \pi/2$ with $\Delta E_{\text{Z},z} = (13 \pm 10)$~$\mu$eV, corresponding to $g^\star_z = 0.2\pm0.2$. For the magnetic field along the $z$-axis $\phi_1 = -\pi/2$ the peak is approximately twice as high and also significantly broadened compared to the $\phi_1 = -\pi$ peak. This indicates that here the Zeeman splitting is too small for the two peaks of the spin-excited and spin-ground state to be resolved. However, the broadened peak can be fitted very well with two peaks that have approximately the height and width of the peak for the spin-ground state measured along the $y$-axis. This provides further confirmation that the broadened peak is indeed a superposition of two separate peaks.
\paragraph{}
For the second measurement $\vec{B}$ always points in a direction perpendicular to the nanowire and is rotated from the $y$ axis at $\phi_2 = 0$ (the same field direction as for $\phi_1 = 0$ in Fig 3(a) to the $x$ axis at $\phi_2 = \pi/2$. Along the $y$ axis, the Zeeman splitting is again $\Delta E_{\text{Z},y} = 154 \pm 5$~$\mu$eV. The Zeeman splitting decreases until it reaches $\Delta E_{\text{Z},x} = 120 \pm 10$~$\mu$eV along the $x$-axis, which corresponds to a $g$-factor of $g^\star_x = 2.1\pm0.2$. In Fig.~\ref{Fig3}(e) two line cuts from Fig.~\ref{Fig3}(b) along the $y$-axis (red curve) and the $x$-axis (blue curve) are presented along with the fitted curves that were used to calculate $\Delta E_{\text{Z}}$.
\paragraph{}
The third rotation plane is the $x$-$z$ plane, with $\phi_3 = 0$ pointing along the $x$-axis, and $\phi_3 = \pi/2$ along the $z$-axis [see Fig.~\ref{Fig3}(c)]. At $\phi_3 = 0$ we measure a spin splitting of $\Delta E_{Z,x} = 117 \pm 10$~$\mu$eV, corresponding to a $g$-factor of $g^\star_x = 2.0\pm0.2$. Rotation of $\vec{B}$ towards the $z$ axis  again results in a Zeeman splitting of $\Delta E_{Z,z} = 17 \pm 10$~$\mu$eV, corresponding to $g^\star_{z} = 0.3\pm0.2$.
\paragraph{}
Combining the three rotation experiments, the Zeeman splitting along each of the $x$, $y$, and $z$ axes is measured twice with consistent values across experiments for the effective $g$-factor. Thus we can summarize our findings in Fig.~\ref{Fig3}(f).
\paragraph{}
Let us now compare our findings with experimental and theoretical results from the literature. An anisotropy of the effective $g$-factor has been measured in other systems like Si nanowire metal-oxide-semiconductor field-effect transistors (MOSFETs) ($g^\star_{\text{max}}/g^\star_{\text{min}} \approx 1.7$) \citep{Voisin2015}, InAs nanowires ($g^\star_{\text{max}}/g^\star_{\text{min}} \approx 1.3$) \citep{Schroer2011}, and InSb nanowires ($g^\star_{\text{max}}/g^\star_{\text{min}} \approx 1.5$) \citep{Nadj-Perge2012a}, all an order of magnitude smaller than our findings of $g^\star_{\text{max}}/g^\star_{\text{min}} \approx 13$ for rotations with respect to the nanowire axis. Also self-assembled SiGe islands on Si have been used for studies on the anisotropy of the effective $g$-factor ($g^\star_{\text{max}}/g^\star_{\text{min}} \approx 5$) \citep{Katsaros2010}, a system that is similar to ours, but lacking the one-dimensional confinement as well as the pronounced strain of our nanowires. None of the reported anisotropies has been attributed to tunable electric fields, where we find $g^\star_{\text{max}}/g^\star_{\text{min}} \approx 1.4$ for rotation with respect to the $E$-field axis. For Ge/Si core-shell nanowires, Hu \emph{et al.} \cite{Hu2012} reported an effective $g$-factor of $g^\star \approx 1.02$ measured with the $B$-field aligned along the nanowire axis $0 \pm 30^{\circ}$. Roddaro \emph{et al.} \cite{Roddaro2008} have measured $g^\star$ for different transitions ranging from 1.6 to 2.2, $\vec{B}$ was here aligned perpendicular to the nanowire. Both values are consistent with our measurements. 
\paragraph{}
A $g$-factor anisotropy can in principle be related to the crystal direction \citep{Nowak2011}. In our probably $<$110$>$-oriented device, we have observed a different anisotropy for holes states most probably originating from higher subbands, while we have observed qualitatively the same anisotropy in a second device tuned to the few-hole regime. Therefore we have strong evidence that the $g$-factor anisotropy observed here is rather related to an electric-field induced mixing between the lowest-lying subbands as discussed in the following paragraph.
\paragraph{}
Maier \emph{et al.} \cite{Maier2013} theoretically investigated the $g$-factor in Ge-Si core-shell nanowires. They assumed elongated quantum dots, i.e. $r_{\text{core}} \ll l_{\text{dot}}$, which is very well fulfilled in our device, where $r_{\text{core}} \approx 8$~nm and $l_{\text{dot}} \approx 150$~nm. They predicted the $g$-factor to be highly anisotropic, with $g$ indeed being quenched along the nanowire axis, and a maximum $g$-factor perpendicular to the nanowire. This is in excellent agreement with our measurements. Moreover, they predicted a lower $g$-factor at finite electric fields. In particular, their calculations showed a more effective diminishment for $\vec{B} \parallel \vec{E}$ than for $\vec{B} \perp \vec{E}$. Also this agrees well with our findings. Maier \emph{et al.} show that this tunability of the $g$-factor with electric fields is caused by the effective coupling of different subbands through these electric fields and the mixed heavy-hole light-hole nature of the individual subbands. The combination leads to a very pronounced spin-orbit interaction (SOI) introduced as the `direct Rashba spin-orbit interaction' \cite{Kloeffel2011a} because of its resemblance of the standard Rashba SOI and the fact that it is a leading-order process not suppressed by the band gap and thus expected to be 10-100 times stronger than the standard Rashba SOI for geometries similar to our device. 
\paragraph{}
Also quantitatively our measurements agree very well with the predictions regarding the $g$-factor quenching along the nanowire axis. The $g$-factor suppression by the electric field is less pronounced than the calculations. This can be explained by differences in the exact geometry of the wires, the quantum dot not being in the single-hole regime, and the fact that our device is operated at significantly higher electric fields than assumed by Maier \emph{et al}.
\paragraph{}
Let us now briefly discuss the implications of these results for quantum computation applications. A main obstacle for spin-based qubits is the fast coherent manipulation of the spin state. In principle this can be done with a pulsed high-frequency (HF) magnetic field, but this is technologically very challenging. Our results imply that it is not only possible to use pulsed HF electric fields as also used in other systems with significant SOI \citep{Nowack2008,Nadj-Perge2010b}, but with continuous HF electric fields while tuning the qubit in and out of resonance by changing the static electric field, \eg, through a combination of top and bottom gates.

\section{Conclusion}

In conclusion, we have demonstrated control over the hole occupancy in a Ge-Si core-shell nanowire quantum dot over 60 charge transitions. The effective $g$-factor has been found to be highly anisotropic with respect not only to the nanowire axis but also the electric-field direction. In particular we have found excellent qualitative agreement between our measurements and theoretical calculations \cite{Maier2013}. This opens the way to controlled manipulation of the spin-state with a continuous high-frequency electric field, a major technological advancement.

\begin{acknowledgments}
We thank Christoph Kloeffel, Daniel Loss, and Wilfred van der Wiel for fruitful discussions and careful reading of the manuscript. We acknowledge technical support by Sergey Amitonov, Paul-Christiaan Spruijtenburg, and Hans Mertens. F.A.Z. acknowledges financial support through the EC Seventh Framework Programme (FP7-ICT) initiative under Project SiAM No. 610637, and from the Foundation for Fundamental Research on Matter (FOM), which is part of the Netherlands Organization for Scientific Research (NWO). E.P.A.M.B. acknowledges financial support through the EC Seventh Framework Programme (FP7-ICT) initiative under Project SiSpin No. 323841.
\end{acknowledgments}

\end{document}